\documentclass[12pt]{article}
\usepackage{graphicx}
\usepackage{psfrag}

\newcommand{\gsim}{\lower.7ex\hbox{$
\;\stackrel{\textstyle>}{\sim}\;$}}
\newcommand{\lsim}{\lower.7ex\hbox{$
\;\stackrel{\textstyle<}{\sim}\;$}}

\def\beq{\begin{eqnarray}}
\def\eeq{\end{eqnarray}}
\def\M{M_{(5)}}

\begin{document}

\begin{flushright}
NYU-TH/01/06/02 \\
TPI-MINN-01/31\\
UMN-TH-2015 \\
astro-ph/0106449\\
\end{flushright}

\vskip 1cm
\begin{center}
{\Large \bf Comments on ``A Supernova Brane Scan''
}\vskip 2cm
{C\'edric Deffayet$^1$, Gia Dvali$^1$, and Gregory Gabadadze$^2$}

\vskip 1cm 
{\it $^1$Department of Physics, New York University, New York, NY 10003\\
$^2$Theoretical Physics Institute, University of Mineesota, Minneapolis, 
MN 55455}\\
\end{center}

\vspace{0.9cm}
\begin{center}
{\bf Abstract}
\end{center}
Recently we  proposed that the acceleration of the Universe
could be due to gravity leaking to extra dimensions.
This scenario gives an alternative to the cosmological constant
or quintessence. In [astro-ph/0106274] the authors claimed  
that this proposal is strongly disfavored 
if not ruled out by the existing data  on 
Supernovae, Cosmic Microwave Background and 
clusters.  We show that the work [astro-ph/0106274] contains
incorrect statements in the theoretical part,
and, moreover, we argue that the main conclusions  of the work 
concerning the comparison with the data are premature.

\newpage

In Ref. \cite {DDG} we proposed a brane-world scenario
in which the late time acceleration of the Universe 
is due to the fact that gravity at ultra-large distances becomes 
five-dimensional and can leak into the extra fifth dimension.
This proposal is an alternative to the cosmological constant
or quintessence scenarios for the accelerated Universe. It is 
based on a brane-world model of Ref. \cite{DGP}, and on a 
subclass of cosmological solutions found in Ref. \cite {Cedric}.

We analyzed qualitatively in \cite {DDG} the Supernovae and 
Cosmic Microwave Background data.
The analysis in \cite{DDG} indicated that the existing data cannot  
discriminate between our model and standard 
cosmology, but the results of further precision measurements 
might do so. The next step is of course the 
confrontation with the real data.  

Recently, the authors of Ref. 
\cite{Avelino} claimed that the model \cite{DGP}, and the 
proposal \cite {DDG} is strongly disfavored if not 
ruled out completely by
the existing data on Type Ia Supernovae, 
Cosmic Microwave Background and clusters.
The purpose of the present note is to 
discuss the work of Ref. \cite {Avelino}. In  
particular, we will argue that these claims are premature.

First we briefly review   
the main features of the model \cite {DGP} and 
then turn to our proposal \cite {DDG}.

The model of Ref. \cite {DGP} 
describes a four-dimensional (4D) world which is 
confined to a hyper-surface (a ``brane'') in five-dimensional 
(5D) space with infinite flat extra dimension. 
The electromagnetic, weak and strong interactions
are ``stuck'' to the brane hyper-surface, 
however gravity is an exception and can propagate 
everywhere in the whole 5D space. As a result,
the gravitational interactions are different 
in our 4D world: Gravity behaves in
a conventional Newtonian way  at observable distances, 
however, the interactions are 
modified at ultra-large cosmological distances where 
they become five-dimensional.

From the phenomenological standpoint, 
there is one free parameter in the model of
\cite {DGP}, that is the 5D gravity scale  $\M$.
The latter controls the crossover distance $r_c$ 
between the conventional 4D and unconventional 5D regimes:
\beq
r_c ~=~{M_P^2 \over 2 M_{(5)}^3}~, 
\label{rc}
\eeq
where $M_P\sim 10^{18}$ GeV is the reduced 4D
Planck mass.  For distances $r$ smaller than $r_c$ 
gravity is conventional and gives the $1/r^2$ Newton
law. However, for distances bigger than $r_c$ 
this breaks down and gravity enters the 5D regime.  
As was pointed out in Ref.  
\cite {DGKN1}, and shown in detail in Ref. 
\cite {DGKN2} the high-energy particle processes,
astrophysical constraints 
and the results of sub-millimeter gravitational measurements
put the lower bound on $\M$, that is 
$\M~\gsim~10^{-3}~{\rm eV}~$. On the other hand, 
the requirement that the cosmological evolution
be conventional four-dimensional until the 
late times requires that $ \M~\lsim~ 10~{\rm MeV}~$. 
Thus, we have the following range for the parameter $\M$:
\beq
10^{-3}~{\rm eV}~\lsim \M~\lsim~ 10^{7}~{\rm eV}~.
\label{M}
\eeq
Using (\ref {rc}) 
these constraints on $\M$ translate roughly into the following  
admissible range for the crossover distance $r_c$:
\beq
10^{29}~{\rm cm}~\lsim~r_c~\lsim~10^{59}~{\rm cm}~.
\label{rc1}
\eeq
Thus, the lowest possible value of $r_c$ is of the order of 
present-day horizon size. 

Different values of $r_c$ are 
interesting from a different perspective.
A model with $r_c \sim 10^{59}~{\rm cm}$, 
i.e., with $\M\sim 10^{-3}~{\rm eV}~$, 
predicts a deviation from the Newtonian gravity at 
very small sub-millimeter distances (see detailed discussions in
\cite{DGKN2})  where the Newtonian gravity is just being tested 
experimentally \cite {Adel}. 
For this value of $r_c$,
in order to obtain the accelerated Universe  
one has to add a cosmological constant or 
a quintessence field on a brane.     
In this case, the model of \cite {DGP} cannot be distinguished 
from the conventional 4D scenario
on the basis of cosmological arguments 
(but it can be distinguished in
sub-millimeter measurements of the gravitation law).

From the point of view of cosmology it is  
interesting to consider another range of the 
parameter space in Eqs. (\ref {M}), (\ref {rc}),
that is  $ \M~\sim~ 10^{7}~{\rm eV}~ $, i.e., 
$r_c \sim 10^{29}~{\rm cm}$. As was discussed in Ref. 
\cite {DDG} there are interesting cosmological 
predictions for this value of the parameter $\M$ 
if one uses the self-accelerated cosmological 
solution of Ref. \cite {Cedric}.

After this brief introduction we comment on the 
results of Ref. \cite {Avelino}.

The first comment concerns the statements made 
in the abstract, introduction and conclusions
of \cite {Avelino} claiming that the model of \cite {DGP}
(which the authors of \cite {Avelino} call the DGP model)
is ruled out by  the cosmological data. This statement is false  
since the model of \cite {DGP} can in no way be ruled out by cosmology
as  we just discussed. Indeed, for the 
most  part of the parameter space in (\ref {M})
the model of \cite {DGP} predicts no changes in conventional
4D cosmology.   
For instance, if $M_{5}\sim 10^{-3} {\rm eV}$, the crossover scale 
$r_c$ is about 30 orders of magnitude
bigger than the present day horizon size and the
cosmology of the model \cite {DGP} is identical for all the 
practical purposes to the conventional 
4D cosmology.

Let us now turn to the proposal of Ref. \cite {DDG}.
There we chose  the lowest possible value of the parameter
$r_c\sim 10^{29}~{\rm cm}$. In this case the modification of gravity 
takes place at the size of the present day horizon. We confined
ourselves to a particular cosmological
solution with self-acceleration which was found in 
Ref. \cite{Cedric} and proposed in \cite {DDG} to use this solution
for the description of the Supernovae data on the acceleration of
the Universe \cite {cc}. It is only this proposal which could  be ruled out
in principle in \cite {Avelino} by the cosmological data.
In other words, cosmological considerations
could rule out the part of the parameter space
of the model \cite {DGP} with $\M \sim 10 $ MeV, but not the model 
of \cite {DGP} itself as claimed in \cite {Avelino}.

Moreover, we will ague below that 
given the state of the current data
the proposal of \cite {DDG} cannot be favored over the conventional 
scenarios, and certainly cannot be  disfavored or 
excluded, in contrast with the conclusions of Ref. \cite {Avelino}. 
We will turn to this momentarily after a minor digression in the 
next paragraph.

Here it is appropriate  to make one more comment on the 
claims of Ref. \cite {Avelino}.
Authors of \cite {Avelino} state that one cannot accommodate on a brane
with infinite extra space the closed ($k=1$) universe 
since the latter  requires a finite volume. 
Although this issue has only an academic 
interest (as the current data supports the 
flat Universe with $k=0$), nevertheless,  for the sake of 
completeness we will comment on this.
The statement of \cite {Avelino} 
is not correct since the $k=1$ solutions can be accommodated on the 
brane along the lines of Refs. \cite {Cedric},\cite {Lue}.
Since this is just a straightforward rewriting of the equations of 
Refs. \cite {Cedric},\cite {Lue} for the $k=1$ case 
we will not repeat them here. It seems that the authors of 
\cite {Avelino}  did not realize  the fact that even if  
the volume of the extra space is infinite,  
this {\it does not mean} that the 3D 
volume of the observable braneworld Universe should 
also be infinite
\footnote{As an additional side comment 
we just briefly mention that the problem of the 
extra polarization of the 5D graviton which was discussed  
in \cite {DGP} and is a point of concern in \cite {Avelino}, 
was recently resolved in \cite {Arkady}.}.

The cosmology of the model considered in \cite{DDG} is 
governed by the following first Friedmann equation
\begin{equation} \label{Frie}
H(z)^2~ =~ H_0^2~ \left\{\Omega_k(1+z)^2~ + ~\left( \sqrt{\Omega_{r_c}} + 
\sqrt{\Omega_{r_c} + \Omega_M (1+z)^3} \right)^2
 \right\}~,
\end{equation}
where $\Omega_{r_c}$ is defined by $\Omega_{r_c} \equiv (2 r_c H_0)^{-2}$ 
and the only energy-momentum content of the universe is 
non relativistic matter (and no cosmological constant). 
The Friedmann equation is thus non standard as is the 
normalization condition
\footnote{Note that $\Omega_M$ and $\Omega_k$ are defined in a conventional
way.} 
\begin{equation}
\Omega_k + \left( \sqrt{\Omega_{r_c}} + 
\sqrt{\Omega_{r_c} + \Omega_M} \right)^2 = 1.
\end{equation}  
The outcome of equation (\ref{Frie}) is that the early cosmology is standard
matter-dominated, whereas the late time 
cosmology exhibits acceleration when the matter density becomes  
small enough. 
This late time behavior enables to 
explain the Supernovae observations \cite{cc} 
without the need of a cosmological constant or quintessence field 
\cite {DDG}.

Let us now turn to the issue of  comparison with  
the data.

As it is easy to see  from the plots presented in 
Ref. \cite{DDG}  one needs a lower value of $\Omega_M$ 
than in standard cosmology in order 
to fit the SN data in our model. 
Our current  estimate  \cite{Raux}, 
which is based on a fit of 54 Supernovae  
(18 nearby and 36 distant, 
taken from second Ref. of \cite{cc}) is for a flat universe 
(as required by CMB observations) 
at one sigma level
\begin{equation} \label{estim1}
\Omega_M = 0.24^{+0.08}_{-0.06}~.
\end{equation}
This should 
be compared with the number quoted in 
\cite{Avelino} obtained for our model by fitting 92 Supernovae 
(at one sigma level)\footnote
{and $\Omega_M = 0.2^{+0.1}_{-0.1}$ at $99 \%$ 
confidence level}:
\begin{equation} \label{estim2}
\Omega_M = 0.2^{+0.05}_{-0.05}.
\end{equation}
It is at least an overstatement 
to claim, as in \cite{Avelino}, that such numbers are enough 
to rule out the model on the basis of  estimates of 
$\Omega_M$ coming from other means than SN or CMB observations.
One can certainly pick a single, least favorable  value
among these estimates to compare it to our model, as it was 
done in \cite{Avelino}, but this cannot be justified by any scientific
means. The value of $\Omega_M$ obtained
from fitting the data which is  quoted in \cite{Avelino} is 
$\Omega_M = 0.35^{+0.07}_{-0.07}$. Even if one is to accept  
this value and (\ref{estim2}) for granted, 
one could  really wonder  
whether a two sigma discrepancy is enough to rule out a model, 
given the state of the art of  measurements of $\Omega_M$.
The estimates  for  $\Omega_M$ from other means than Supernovae data 
(and CMB) is far from being settled with such a good precision to 
rule out (\ref{estim1}) or (\ref{estim2}). Moreover,  
it is well known that the different measurements have 
systematic uncertainties which are very difficult to 
assign  (see for example the discrepancy between the values 
quoted in \cite{Carlberg} and the one obtained 
by X-Ray or SZ measurements quoted in \cite{Avelino}). 
We will not comment on this any longer, 
let a reader decide if numbers such as 
(\ref{estim1}) or (\ref{estim2}) are good enough to 
rule out  a model.

On the other hand, as pointed out  in \cite{DDG},
it is perfectly true that the model can be disproved
with coming data and precision cosmological tests,
and this makes it exciting. 
However, given the status of the current data 
it does not seem reasonable to make 
strong statements such as those 
made in Ref. \cite{Avelino}. 

Some of us are currently working on a more comprehensive 
comparison with the data including the large 
scale structure and CMB results \cite{Prep}.   

\vspace{0.1in}

We would like to thank Pierre Astier, Andrei Gruzinov, 
David Hogg, Jim Peebles, Roman Scoccimarro and  Matias Zaldarriaga  
for useful discussions.


\begin{thebibliography}{99}

\bibitem{DDG}
C.~Deffayet, G.~Dvali and G.~Gabadadze,
[astro-ph/0105068].

\bibitem{DGP} G.~Dvali, G.~Gabadadze and M.~Porrati,
Phys.\ Lett.\  {\bf B485} (2000) 208
[hep-th/0005016].

\bibitem{Cedric}
C.~Deffayet,
Phys.\ Lett.\ B {\bf 502} (2001) 199
[hep-th/0010186].


\bibitem{Avelino}
P.P. Avelino and C.J.A.P. Martins
``A supernova Brane Scan'' [astro-ph/0106274].

\bibitem{DGKN1}
G.~Dvali, G.~Gabadadze, M.~Kolanovic and F.~Nitti,
hep-ph/0102216.

\bibitem{DGKN2} G.~Dvali, G.~Gabadadze, M.~Kolanovic and F.~Nitti,
hep-th/0106058.

\bibitem{Adel}
C.~D.~Hoyle, U.~Schmidt, B.~R.~
Heckel, E.~G.~Adelberger, J.~H.~Gundlach, D.~J.~Kapner and H.~E.~Swanson,
``Sub-millimeter tests of the gravitational inverse-square law: 
A search  for 'large' extra dimensions,''
hep-ph/0011014;\\
J.C. Price, in {\it Proceedings of 
International Symposium on Experimental Gravitational Physics}, ed. 
Michelson, Guangzhou, China (World Scientific, Singapore,1988);\\
J. Long, ``Laboratory Search for Extra-Dimensional Effects in 
Sub-Millimeter
Regime'' Talk given at the International Conference on Physics Beyond Four 
Dimensions, ICTP, Trieste, Italy; July 3-6, (2000);\\
A. Kapitulnik, ``Experimental Tests of Gravity Below 1mm''
Talk given at the International Conference on Physics Beyond Four 
Dimensions, ICTP, Trieste, Italy; July 3-6, (2000)~.


\bibitem{cc} A.G. Riess et al., {\it Astroph. J} 116, 1009 (1998);\\
S. Perlmutter et al., ``Measurements of Omega 
and Lambda from 42 High-Redshift Supernovae", [astro-ph/9812133];\\
A.G. Riess, Talk Given at The Symposium ``{\it The Dark Universe:
Matter, Energy, and Gravity}'' Baltimore, April 2-5, (2001). 

\bibitem{Lue} C.~Deffayet, G.~Dvali, G.~Gabadadze and A.~Lue,
hep-th/0104201.


\bibitem{Arkady} C.~Deffayet, G.~Dvali, G.~Gabadadze and A.~Vainshtein,
hep-th/0106001.

\bibitem{Raux}
J. Raux, private communication to be published.

\bibitem{Carlberg}
R. Carlberg, {\it et al} 1997, ApJ 478, 462; \\
N. Bahcall, {\it et al} 2000, ApJ 541, 1.

\bibitem{Prep} C. Deffayet, S. Landau, J. Raux, M. Zaldarriaga, 
in preparation.



\end{thebibliography}
\end{document}